# Ferroelectric Multiple-Valued Logic Units

I. Lukyanchuk,[1,2,*] E. Zaitseva,[3] V. Levashenko,[3] M. Kvassay,[3]
S. Kondovych,[1] Yu. Tikhonov,[4] L. Baudry,[5] and A. Razumnaya [4]

[1] Laboratoire de Physique de la Matière Condensée, Université de Picardie Jules Verne, 80039 Amiens, France

[2] Landau Institute for Theoretical Physics, Moscow, Russia

[3] Faculty of Informatics and Management Science, University of Zilina, Zilina 01026, Slovakia

[4] Faculty of Physics, Southern Federal University, Rostov-on-Don 344090, Russia

[5] Institute of Electronics, Microelectronics and Nanotechnology (IEMN)-DHS Départment, UMR CNRS 8520, Université des Sciences et Technologies de Lille, 59652 Villeneuve d'Ascq Cedex, France

Employing many-valued logic (MVL) data processing allows to dramatically increase the performance of computing circuits. Here we propose to employ ferroelectrics for the material implementation of MVL units basing on their ability to pin the polarization as a sequence of multi-stable states. Two conceptual ideas are considered. As the first system, we suggest using the strained ferroelectric films that can host the polarization states, allowing the effective field-induced multilevel switching between them. As the second one, we propose to employ the ferroelectric nano-samples that confine the topologically-protected polarization textures which may be used as MVL structural elements. We demonstrate that these systems are suitable for engineering of the 3-, 4- and even 5-level logic units and consider the circuit design for such elements.

**KEYWORDS**: Ferroelectrics, switching, hysteresis, topological structures, multiple-valued logic, vortices, domains, skyrmions.

# 1. Introduction

Permanently growing size and amount of processed information and increasing of the computation speed impose the emergence of new technologies for efficient data treatment. A non-binary architecture of the computing circuits is considered as an alternative to the conventional binary 2-valued logic. Binary elements are based on the principle that a data bit is always either "zero" or "one". However, already in the 1970s, the researchers pointed out the limitations of this technology. The first one is the interconnect problem, related to both on-chip and between-chip connections. The difficulties of placement and routing of the digital logic elements are escalating with an increase of the computation capability per chip. In fact, the area of interconnections may consist of more than 70% of active logic elements [1]. However, the further increase of the on- and off-chip connections faces the mechanical, thermal, and electrical restrictions [2,3].

Another limiting factor is the necessity of increasing the clock speed of switching between different binary states [4] that finally determines the computer performance. During past decades, the clock speed had doubled almost every year. Usually, the limitation of the clock speed is bypassed by packing some cores into a chip, which has resulted in multi-core processors. However, this approach does not greatly improve performance because of the limiting amount of the binary data that need to be transferred.

These factors point on the necessity to design the principally new computing circuits using the *Multiple-Valued Logic* (MVL) architecture. The primary advantage of MVL is the ability to encode more information per variable ("multi-valued bit") [1,5]. However, in this case, one should go beyond the conventional material technology of semiconducting transistor, having only two stable states, "on" or "off". Hence, the development of new MVL computing circuits is based on two inter-related advances, the technological one that includes the development of new non-silicon multi-valued logic gates and the mathematical one that embraces the design of new computation methods and algorithms.

In this work, we suggest to use the ferroelectric materials as a platform for implementation of the MVL elements. The idea is based on the capacity of ferroelectrics to hosts multiple polarization states having nearly the same energy. Forming the logical levels of the MVL unit, these states can be switched by the electric field. The most evident way that we consider in Section 2 is to use the degeneracy of polarization orientation in the pseudo-cubic crystal lattice of perovskite oxides. As a matter of fact, the strained films of pseudo-cubic ferroelectric oxides can host a variety of logically different multilevel hysteresis loops, holding two, three, or even four polarization states [6,7]. Importantly, these logical gates are operational at room temperature, and their logic is tunable by strain and temperature which is important for prospective future "in silico" applications.

Even more fascinating opportunity can be provided by a variety of topological structures of polarization, confined in the nanoscale ferroelectric samples: nanodots, and nanopillars (nanorods). In Section 3 we consider the switching between different types of these structures: domain patterns [8,9], singular vortices [10-12], coreless vortices and skyrmions [13-16] that allows to realize even more lively logical states for MVL elements, opening this the unprecedented horizons for the domain- and topological structures – provided nanoelectronics [17].

The design of the logic operations and MVL circuits will be considered in Section 4. Our findings enable developing a platform for the emergent MVL technology of information processing and target the further challenges posed by needs of quantum and neuromorphic computing.

## 2. Multilevel hysteresis switching in ferroelectrics films

In [7] we have demonstrated that MVL cells can be realized using the substrate-deposited thin films of ferroelectric perovskite oxides. The model ferroelectric material, $PbTiO_3$, having the pseudo-cubic structure is viewed as a promising material since the technology of thin-film deposition of $PbTiO_3$ is fairly well controlled and it can operate at room temperature. Importantly, the polarization states of $PbTiO_3$ films crucially depend on the strain, imposed by the substrate that can have both the tensile and compressive character. Accordingly, the polarization can have either in- or out-of-plane orientations with respect to the plane of the film. The shown in Fig.1a strain-temperature $u_m$-$T$ phase diagram of the substrate-deposited single-domain $PbTiO_3$ films was demonstrated [18] to host three ferroelectric phases: the *c*-phase with the out-of-plane polarization orientation, the *aa*-phase with the in-plane polarization orientation along the face diagonal of the pseudo-cubic lattice cell and the *r*-phase in which polarization is tilted with respect to the film plane and lies in the vertical diagonal sectional plane, formed by the *c*- and *aa*- directions.

The key point here is that staying in one of the thermodynamically stable states, the system can host other metastable states that are the legacy of the phases, stable in other parts of the $u_m$-$T$ phase diagram. Application of the external electric field permits to switch between these states, realizing hence the complicate hysteresis loop with various branches. Depending on the landscape of the energy profile and on the protocol of the field variation one can get the loop of a rather complex structure. Some examples of the hysteretic transitions between *c*- *aa*- and *r*-state having two, three, and four branches are shown in Fig. 1b. (Fig. 1a shows the location of these loops at the $u_m$-$T$ phase diagram.) Panels of Fig. 1b display also the switching maps of the corresponding MVL elements.

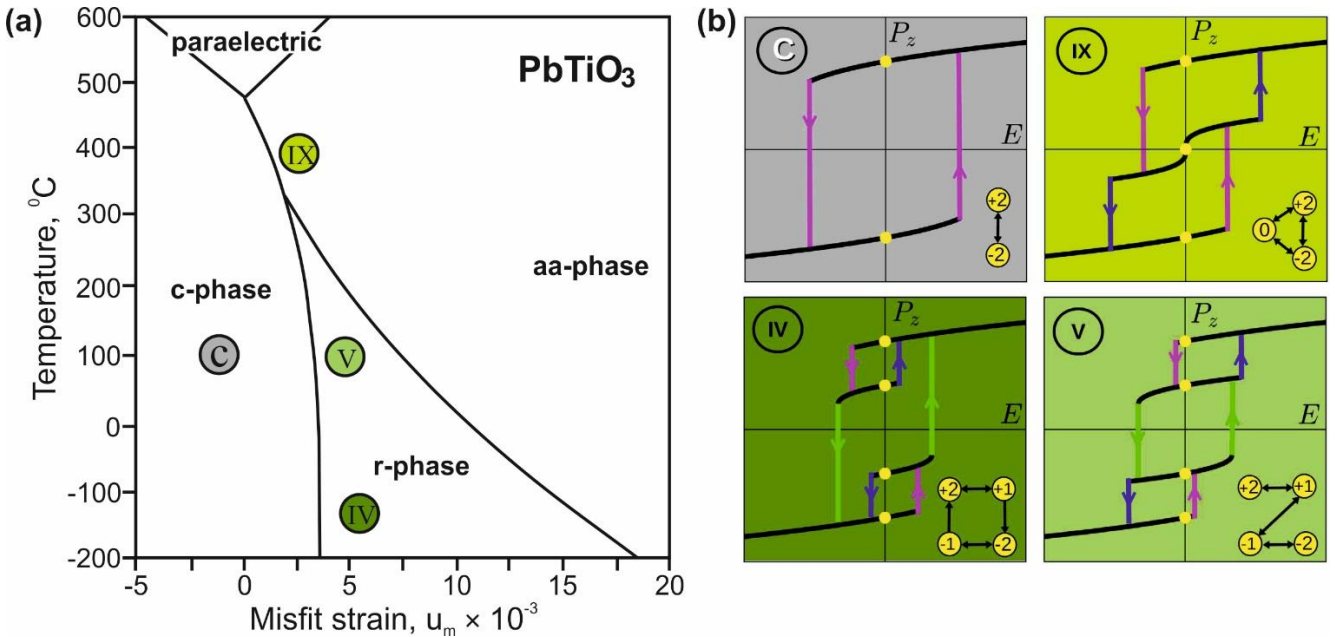

**Figure 1**. **Multilevel hysteresis switching in films of $PbTiO_3$** (**a**) Strain-temperature phase diagram of strained $PbTiO_3$ film [18]. Bubble notations correspond to the location of the hysteresis loops, shown on the panel (b). (**b**) Example of hysteresis loops in $PbTiO_3$ film [7]. Switching topology depends on the location of the loop in panel (a). States "±1" and "±2" correspond to the *r*- and *c*-phases with "up"/"down"-oriented polarization respectively, state "0" correspond to the *aa*-phase. (C) Usual 2-level binary systems, realizing in *c*-phase. (IX) The 3-level loop in *aa*- phase where each state can be reached from all others (IV,V) Examples of 4-level loops in *r*-phase. Insets show the logical switching map of corresponding MVL elements.

Panel C of Fig, 1b shows the 2-branch hysteresis loop, realizing the conventional 2-bit logic operation. This standard situation is realized in the *c*-phase. The *aa*-phase in the vicinity of the first-order transition from *c*-phase hosts a stable *aa*-state and also two metastable *c*-states having the "up" and "down" out-of-plane-oriented polarization. The corresponding 3-branch hysteresis loop (panel IX) allows for the stack-wise access to all the logical levels, realizing hence the 3-level MVL unit. We consider the operational modes for such element in Section 4 in more detail. Finally, the *r*-phase has two "up" and "down" out-of-plane-oriented polarization components and also two metastable states corresponding to the "up"- and "down"-oriented polarization states of the *c*-phase. The corresponding hysteresis with 4 branches can have far more complex logical structure. Two examples of the 4-branch loops and corresponding switching maps of 4-level MVLs are shown at panels IV and V. A more exhaustive description of all the possibilities is given in Ref. [7].

## 3. Switching of confined topological structures

Confinement that imposes additional symmetries on the system can stabilize exotic topological states bringing novel functionalities, which do not exist in bulk materials [19,20]. Topological structures in nano-size ferroelectric samples [21], nanodots, and nanopillars (nanorods) (Fig. 2a) are of special interest for applications because they can be relatively easily controlled and manipulated by electric fields. Ferroelectric topological excitations can be reduced in size to atomic scales, in addition to being tunable through lattice strain. The depolarizing charge associated with the topological excitation permits coupling with the incident electromagnetic field and mutual electrostatic cross-talk.

Importantly, confined topological structures can be manipulated and switched by electric fields with a critical threshold field $E_c <$ 10 mV/nm and characteristic read/write times < 50 ps. Variety of different topologically-protected polarization textures with nearly the same energy arising in the same sample permits the field-induced switching between them, realizing hence the multivalued hysteresis loops, similar to those, described in the previous Section.

The topological class of the confined polarization texture depends on the degeneracy and anisotropy of configuration space of the order parameter, which is the vector of polarization. Figure 2 summarizes the state-of-the-art of the study of topological excitations in ferroelectrics and the field-induced switching between them. They are conventionally classified according to the uniaxial anisotropy of the system, imposed by strains. For simplicity, we do not consider the effect of the higher-symmetry cubic anisotropy, appropriate for the perovskite oxide ferroelectrics.

(i) Uniaxial easy-axis anisotropy, according to magnetic terminology (Fig. 2b). Polarization has only two, equilibrium orientations "up" and "down" with two-point degeneracy $Z_2$ in the configuration space. The Kittel domain structure is formed to minimize the depolarization energy, similar to the ferroelectric thin films and superlattices [22-24]. However, in this case, the confined domain configuration has more elaborated geometry. A wealth of switching paths between them is possible [9]. We show, as an example, the hysteresis loop with 5 branches that realizes the 5-level MVL cell.

(ii) Uniaxial easy-plane anisotropy (Fig. 2c). The in-plane orientation of polarization has the circular degeneracy $S_2$ in configuration space. Calculations, based on the Landau-Ginzburg formalism coupled with the electrostatic equations show that a variety of the vortex-like solutions can have approximately the same minimal energy [11]. The residual degeneracy of the system is characterized by the clock- and counter-clockwise polarisation rotation. The vortices are formed to vanish the depolarization energy that would be huge in case of the uniform polarization. The field, applied perpendicular to the vortex plane results in the 3-branch hysteresis loop that switches between the vortex-states and the "up"- and "down"-polarized monodomain states. Hence, the 3-level MVL cell is realized.

(iii) Almost-isotropic system with spherical $S_3$ degeneracy of the order parameter (Fig. 2d). The exotic skyrmion topological excitation having the structure of vortices with the non-singular core can be confined in cylindrical nanodot [13]. Besides orientation of polarization "up" and "down" inside of the central core region the skyrmions are characterized by the left-hand and right-hand chirality, hence are optically active. Switching of the skyrmions by electric field [14] results in the 4-branch hysteresis loop with the "up"- and "down"-oriented skyrmion states and with "up"- and "down"-oriented uniformly polarized states. This loop corresponds to the 4-level MVL cell.

Many other confined topological structures in ferroelectrics, including the recently reported polarization textures in spherical nanodots [25] and two-domain structure in ferroelectric nano-capacitors with negative capacitance [26] can also be thought as a platform for MVL cells.

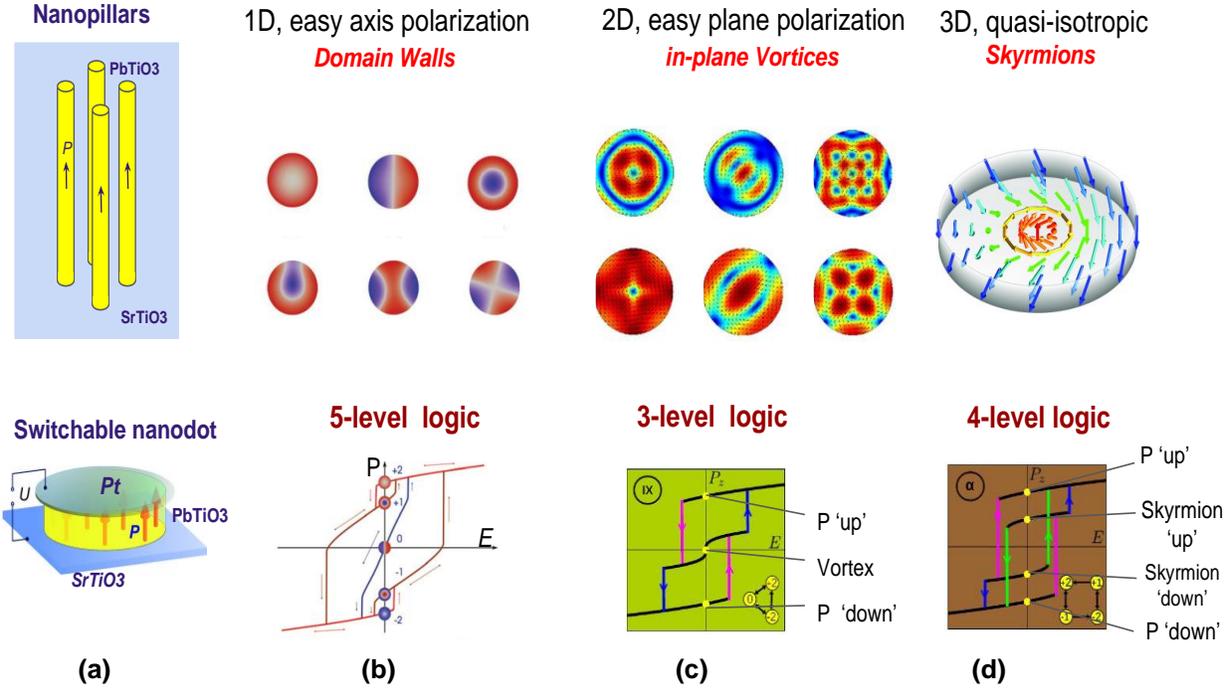

**Figure 2. Multilevel switching between polarization topological structures in nano-samples.** (**a**) Sketch of the PbTiO$_3$ nanopillars embedded in SrTiO$_3$ matrix and of PbTiO$_3$ nanodot deposited on SrTiO$_3$-substrate. (**b**) In case of the uniaxial easy-axis anisotropy, the "up"- and "down"-oriented polarization domains are formed [9] Top view of polarization distribution in domains and example of 5-level switching loop between them are shown. Red and blue colors denote the "up"- and "down"- oriented polarization. (**c**) In the case of uniaxial easy-axis anisotropy, the in-plane polarization vortices are formed [11]. Multiple vortex states are shown. The blue color corresponds to the singular vortex cores where polarization vanishes. The 3-level switching occurs between "up"- and "down"-oriented polarized monodomain states and the in-plane vortex state. (**d**). In case of weak anisotropy, the coreless vortices - skyrmions with "up"- and "down"-oriented cores can be formed in cylindrical nanodot [14]. The switching hysteresis in this case have 4 branches.

## 4. Multi-valued logic circuit design

There are different techniques for logic circuits design based on MVL elements [27-29]. One of these techniques is *Programmable Logic Arrays* (PLA) development that permits to implement any logic function (combinational logic circuits) [28]. PLA is a kind of programmable logic device that has two parts (Fig. 3): memory (M) and a logic block (L). The ferroelectrics elements can be used for the memory block implementation according to the technological aspect.

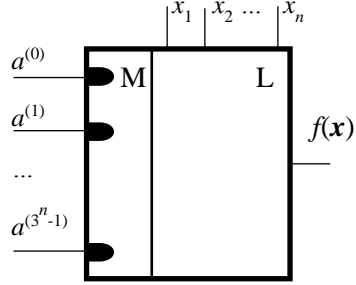

**Figure 3.** Structure of the proposed PLA for implementation of MVL function.

The regular structure of PLA causes the use of canonical and regular mathematical representation of MVL function implemented by this circuit. One of possible representations is generalised Reed-Muller expansion (GRME). The ferroelectrics technologies considered above allows implementing of 3-level switching elements. The mathematical interpretation of this elements in MVL is possible by the 3-valued function $f(x) = f(x_1, x_2, \ldots, x_n)$ of $n$ variables $x_1, x_2, \ldots, x_n$. This function is defined for set $m = \{0, 1, 2\}$ as mapping $\{0, 1, 2\}^n \to \{0, 1, 2\}$. The GRME of 3-valued function $f(x)$ on $n$ variables is determined by the following equation [30-31]:

$$A(x) = \sum_{i=0}^{3^n-1} a^{(i)} x_1^{i_1} x_2^{i_2} \ldots x_n^{i_n} \quad (mod\ 3) \tag{1}$$

were $a^{(i)}$ are the coefficients, $a^{(i)} \in \{0, 1, 2\}$; $i_j$ is the $j$-th digit of 3-valued representation of parameter $i$, $i = (i_1, i_2, \ldots, i_n)_3$ and $j = 1, \ldots, n$. The GRME of 3-function on one variable according to (1) is:

$$A(x) = \sum_{i=1}^{2} a^{(i)} x^i = a^{(0)} + a^{(1)} x + a^{(2)} x^2 = a^{(0)} + (a^{(1)} + a^{(2)} x) x \quad (mod\ 3). \tag{2}$$

The PLA for a 3-valued function based on GRME representation has $3^n$ inputs. These inputs are the memory's inputs to program the GRME coefficients values. The change of these inputs values (re-programming of GRME coefficient) results in new 3-valued function (new circuits based on PLA). The coefficients are read and transmitted to the logic block. The logic block has a set of $n$ external inputs for variables and $3^n$ inputs from the memory block. The logic block has a number of sum-product's homogeneous sub-blocks which are linked together to give output.

Let us consider this PLA for 3-valued functions of tree variable ($n = 3$). The detailed structure of the PLA for 3-valued function on 3 variables is shown in Fig. 4. This PLA consists of memory and logic blocks and has 3 inputs for function's variables $x_1$, $x_2$ and $x_3$ and 27 inputs to programme the coefficients of GRME. The memory block includes 27 memory cells that can be implemented based on ferroelectric technology. The logic block consists of 3 levels. The first level includes 9 homogeneous sub-blocks that are connected to 3 sub-blocks of the second level. The third level consists of one sub-block. All sub-block are homogeneous and each of them has 3 inputs from the previous level, 1 input for variable and 1 output. The logical structure of the sub-block is caused by the GRME for one variable (2). The sub-block is the parallel structure with 3 modulo adders and 2 modulo multipliers that implement the calculation of (2). The modulo adder and multiplier can be elaborated based on different technologies, for example, these devices based on CMOS technology is considered in [32].

The logic block for the 3-valued function of $n$ variable is homogeneous recurrent and parallel. It consists of $n$ level. The $s$-th level ($s = 1, \ldots, n$) is connected to the ($n-s+1$)-th variable and includes ($3^{n-s}$) homogeneous sub-blocks each of them calculates GRME on one variable (2). The outputs of blocks of the $s$-th level are inputs of blocks of the ($s+1$)-th level. The first level is connected to the memory of PLA to read the coefficients.

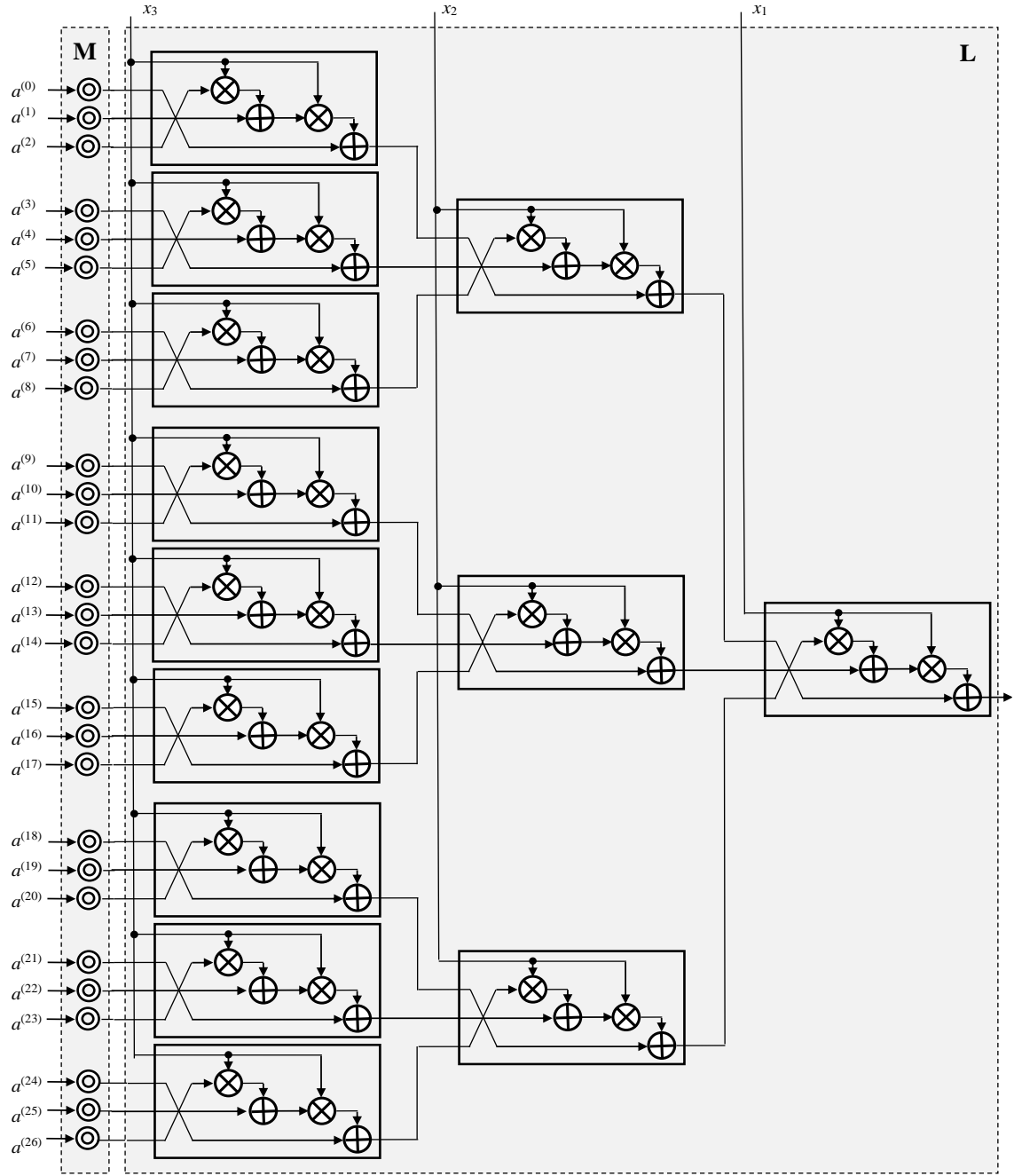

**Figure 4.** The detailed structure of the PLA for implementation of 3-valued function on 3 variables

## Acknowledgments

This work was supported by the French-Slovak bilateral collaborative program PHC-STEFANIK, by the Slovak Research and Development Agency under the contract No. SK-FR-2017-0003 and by the H2020-RISE-ENGIMA action.

## References

1. S. L. Hurst, Multiple-Valued Logic. Its Status and its Future. *IEEE Trans on Computers* **33** (12), 1160 – 1179 (1984).
2. P. Beckett, Towards a Reconfigurable Nanocomputer Platform, *7th ACSAC Proc.*, 141-150 (2002). (P. Beckett, Towards a Reconfigurable Nanocomputer Platform, *Australian Computer Science Communications* **24**, 141-150 (2002).
3. C. Vudadha, and M.B. Srinivas, Design Methodologies for Ternary Logic Circuits, *48th IEEE Int. Symp. on Multiple-Valued Logic Proc.*, 192 – 197 (2018).
4. B. Choi, and K. Shukla, Multi-Valued Logic Circuit Design and Implementation, *Int. Journal of Electronics and Electrical Engineering* **3** (4), 256 – 262 (2015).
5. A. P. Surhonne, D. Bhattacharjee, and A. Chattopadhyay, Synthesis of Multi-Valued Literal using Łukasiewicz logic, *48th IEEE Int. Symp. on Multiple-Valued Logic Proc.*, 204 – 209 (2018).
6. L. Baudry, I. A. Luk'yanchuk, and A. Razumnaya, Dynamics of Field Induced Polarization Reversal in Strained Perovskite Ferroelectric Films with c-oriented Polarization, *Phys. Rev. B* **91**, 144110 (2015).
7. L. Baudry, I. Lukyanchuk, and V. M. Vinokur, Ferroelectric symmetry-protected multibit memory cell, *Sci. Rep.* **7** 1–7 (2017).
8. I. Lukyanchuk, P. Sharma, T. Nakajima, S. Okamura, J.F. Scott, and A. Gruverman, High-Symmetry Polarization Domains in Low-Symmetry Ferroelectrics, *Nano Lett.* **14**, 6931 (2014).
9. P.-W. Martelli, S. M. Mefire, and I. Luk'yanchuk, Multidomain switching in the ferroelectric nanodots, *Europhys. Lett.* **111**, 50001 (2015).
10. I. I. Naumov, L. Bellaiche and H. Fu, Unusual phase transitions in ferroelectric nanodisks and nanorods, *Nature* **432**, 737 (2004)
11. L. Lahoche, I. Luk'yanchuk, and G. Pascoli, Stability of vortex phases in ferroelectric easy-planes nano-cylinders, *Integrated Ferroelectrics* **99**, 60 (2008).
12. A. K. Yadav, C. T. Nelson, S. L. Hsu, Z. Hong, J. D. Clarkson, C. M. Schlepütz, A. R. Damodaran, P. Shafer, E. Arenholz, L. R. Dedon, D. Chen, A. Vishwanath, A. M. Minor, L. Q. Chen, J. F. Scott, L. W. Martin, and R. Ramesh, *Nature* **530**, 198 (2016).
13. L. Baudry, A. Sene, I. Luk'yanchuk, and L. Lahoche, Vortex state in thin films of multicomponent ferroelectrics, *Thin Solid Films* **519**, 5808 (2011).
14. L. Baudry, A. Sené, I.A. Luk'yanchuk, L. Lahoche, and J.F. Scott, Polarization vortex domains induced by switching electric field in ferroelectric films with circular electrodes, *Phys. Rev*. B **90**, 024102 (2014).
15. Y. Nahas, S. Prokhorenko, L. Louis, Z. Gui, I. Kornev, and L. Bellaiche, Discovery of stable skyrmionic state in ferroelectric nanocomposites, *Nat. Comm.* **6**, 1 (2015).
16. M.A.P. Gonçalves, C. Escorihuela-Sayalero, P. García-Fernández, J. Junquera and Jorge Íñiguez, Theoretical guidelines to create and tune electric skyrmions, *Preprint arXiv*:1806.01617 (2018).
17. G. Catalan, J. Seidel, R. Ramesh, and J. F. Scott, Domain wall nanoelectronics, *Rev. Mod. Phys*. **84**, 119 (2012).
18. N. A. Pertsev, A. G. Zembilgotov, and A. K. Tagantsev, Effect of Mechanical Boundary Conditions on Phase Diagrams of Epitaxial Ferroelectric Thin Films, *Phys. Rev. Lett.*, **80**, 1988 (1998).
19. N. D. Mermin, The topological theory of defects in ordered media, *Rev. Mod. Phys*. **51**, 591 (1979).